\documentclass{article}
\usepackage{spconf,amsmath,graphicx,amsfonts,multirow,booktabs,float,threeparttable,lipsum,dsfont}

\def\x{{\mathbf{x}}}
\def\z{{\mathbf{z}}}
\def\C{{\mathbf{C}}}

\title{an iterative framework for self-supervised \\deep speaker representation learning}
\name{Danwei Cai$^{\star}$, Weiqing Wang$^{\star}$, Ming Li$^{\star \dagger}$}
\address{$^{\star}$Department of Electrical and Computer Engineering, Duke University, Durham, USA\\$^{\dagger}$Data Science Research Center, Duke Kunshan University, Kunshan, China \\\texttt{ming.li369@duke.edu}}

\begin{document}

\maketitle

\begin{abstract}
In this paper, we propose an iterative framework for self-supervised speaker representation learning based on a deep neural network (DNN).
The framework starts with training a self-supervision speaker embedding network by maximizing agreement between different segments within an utterance via a contrastive loss.
Taking advantage of DNN's ability to learn from data with label noise, we propose to cluster the speaker embedding obtained from the previous speaker network and use the subsequent class assignments as pseudo labels to train a new DNN.
Moreover, we iteratively train the speaker network with pseudo labels generated from the previous step to bootstrap the discriminative power of a DNN.
Speaker verification experiments are conducted on the VoxCeleb dataset. The results show that our proposed iterative self-supervised learning framework outperformed previous works using self-supervision. The speaker network after 5 iterations obtains a 61\% performance gain over the speaker embedding model trained with contrastive loss.

\end{abstract}

\section{Introduction}
Speaker recognition refers to identify or verify a claimed speaker by analyzing the given speech from that speaker.
Over the past few years, supervised deep learning methods greatly improve the performance of speaker recognition system \cite{snyder_x-vectors:_2018, cai_exploring_2018,chung2020defence}.
These methods require large-scale datasets to learn discriminative speaker representations.
However, manually annotating speaker labels for a large scale dataset may sometimes be expensive and problematic.
On the other hand, there are vast numbers of unlabeled speech data that can be used for training DNNs.
With self-supervision methods, deep learning can automate the labeling process and benefit from massive amounts of data.

In visual representation learning, most self-supervised methods fall into two classes: generative or discriminative. 
The generative approaches directly model the pixels of input images and do reconstruction \cite{colorization,latent2image}. 
However, learning a full generative model in pixel-level may be computationally expensive for representation learning.
Recently, discriminative approaches based on contrastive learning emerge and show promising results in visual representation learning \cite{cpc1,moco,chen_simple_2020}.

In self-supervised speaker representation learning, Stafylakis \textit{et al.} \cite{stafylakis_self-supervised_2019} propose a generative method to learn speaker embedding with the help of a  phone decoder network. Ravanelli \textit{et al.} \cite{ravanelli_learning_2019} propose to learn speaker representation by maximizing the mutual information between two speech segments within the same utterance. Discriminative approaches based on contrastive learning have recently shown promising results \cite{inoue_semi-supervised_2020, huh_augmentation_2020}. In particular, Huh \textit{et al.}~\cite{huh_augmentation_2020} propose to use a strong data augmentation strategy to improve the generalizability of the learned speaker representation and apply augmentation adversarial training to remove the channel noise caused by the augmentation applied. Audio-visual application is also becoming increasingly popular in self-supervised representation learning. It take advantage of this multi-modal information, i.e., images and sound, of the video data to learn multi-modality based identity representation \cite{nagrani_disentangled_2020,chung_seeing_2020}.

When learning speaker representation without supervision, contrastive learning based methods assume that speech segments from different utterances belong to different speakers. This assumption naturally introduces label error when training the network. Even if the network learns some discriminative information about speaker identities, this label error might drive the speaker network to learn the opposite. In this paper, we propose an iterative, self-evolving framework that learns to avoid this kind of label error. It starts with training a speaker embedding network using contrastive self-supervised learning and generates pseudo labels of the training data with this network. The framework then iteratively trains the speaker network with the pseudo labels and generates new labels using the new converged network. The idea behind the proposed framework is simple: to take advantage of the DNN's ability to learn from data with label noise and bootstrap its discriminative power. This idea is similar to the deep clustering method \cite{song_learning_2020} in unsupervised visual features learning, which performs clustering every few epochs. In this paper, we perform clustering whenever the network is converged.

This work is partly motivated by our previous study on unsupervised speaker recognition \cite{liu_iterative_2014}, which is based on the traditional i-vector/PLDA pipeline. It iteratively optimizes the probabilistic linear discriminant analysis (PLDA) scoring backend with the fixed speaker representation of i-vectors. In this paper, the speaker representation is learned discriminatively using the generated pseudo labels at each iteration. 

\section{Methods}
This section describes the proposed iterative framework for self-supervised speaker embedding learning.
\begin{itemize}
    \item Step 1 (initial round): Train a speaker embedding network with contrastive self-supervised learning.
    \item Step 2:  With the previous speaker embedding network, extract speaker embeddings for the whole training data. Perform a clustering algorithm on the embeddings to generate pseudo labels.
    \item Step 3: Train the speaker embedding network with a classification layer and cross-entropy loss using the generated pseudo labels.
    \item Repeat step 2 and step 3 with limited rounds. Use the last speaker embedding network as the final model.
\end{itemize}

\subsection{Contrastive Self-supervised Learning}
We employ the contrastive self-supervised learning (CSL) framework similar to the framework in \cite{chen_simple_2020, falcon_framework_2020}.
Let $\mathcal{D} = \{\x_1, \cdots, \x_N\}$ be an unlabeled dataset with $N$ utterances, CSL assumes that each data sample defines its own class.
During training, we randomly sample a mini-batch $\mathcal{B} = \{\x_1, \cdots, \x_M\}$ of $M$ utterances from $\mathcal{D}$.
For each utterance $\x_i$, two segments $\x_{i,1}, \x_{i,2}$ are randomly taken and considered as a positive pair. This produces $2M$ data samples in total.
Stochastic data augmentation is then performed on these data samples and the speaker embedding $\z_{i,j}$ are extracted as
\begin{equation}
    \z_{i,j} = f_\theta (\mathrm{aug}(\x_{i,j})), \ j\in\{1,2\}
\end{equation}
where $f_\theta(\cdot)$ is the speaker embedding network with parameters $\theta$, and
$\mathrm{aug}$ represents the stochastic data augmentation process: a reverberation noise and (or) additive background noise are (is) added to the clean segment.
We adapt the contrastive loss from SimCLR \cite{chen_simple_2020} as:
\begin{equation}
\mathcal{L}_\mathrm{CSL} = \frac{1}{2M} \sum_{i=1}^M (l_{i,1} + l_{i,2})
\end{equation}
\begin{equation}
l_{i,j} = - \log \frac{\exp(\cos(\z_{i,1}, \z_{i,2})/\tau)}{\sum_{k=1}^M\sum_{l=1}^2 \mathds{1}_{\substack{k \neq i\\l \neq j}}\exp(\cos(\z_{i,j}, \z_{k,l})/\tau)}
\label{equation:csl}
\end{equation}
where $\mathds{1}$ is an indicator function evaluating $1$ when $k \neq i$ and  $l \neq j$, $\cos$ denotes the cosine similarity and $\tau$ is a temperature parameter to scale the similarity scores.
$l_{i,j}$ can be interpreted as the loss for \textit{\textbf{anchor}} feature $\z_{i,j}$.
It computes positive score for \textit{\textbf{positive}} feature $\z_{i,(j+1)\mathrm{mod}2}$ and negative scores across all $2(M-1)$ \textit{\textbf{negative}} pairs $\{\z_{k,j} | k\neq i, j=1,2\}$.

\subsection{Generating Pseudo Labels by Clustering}

\subsubsection{\textit{k}-means clustering}
Given the speaker embeddings of the training data, we employ a clustering algorithm to generate cluster assignments. In this paper, we use the well-know \textit{k}-means because of its simplicity, fast-speed and capability with large dataset. Let the speaker embedding in $d$-dimensional feature space $\z\in\mathbb{R}^d$, \textit{k}-means learns a centroid matrix $\C\in \mathbb{R}^{d\times k}$ and the cluster assignment $y_i \in \{1,\cdots,k\}$ for each speaker embedding $\z_i$ with the following learning objectives
\begin{equation}
    \min_{\C}\frac{1}{N}\sum_{i=1}^{N}\min_{y_i} \| \z_i -\C_{y_i}\|^2_2
\end{equation}
where $\C_{y_i}$ is the $y_i^\text{th}$ column of the centroid matrix $\C$. The optimal assignments $\{y_1, \cdots, y_N\}$ are used as pseudo labels.

\subsubsection{Purifying pseudo labels}
One problem with the generated pseudo labels is the massive label noise. To mitigate this problem, we apply the following simple steps to purify the generated pseudo labels.
(\textit{a}) By defining the clustering confidence as $-\|\z_i-\C_{y_i}\|_2^2$ for each speaker embedding $\z_i$, we filter out $p$ portion of the remaining data with least clustering confidence.
(\textit{b}) To further reduce the possibility that one actual speaker appears in several pseudo clusters, we only keep the pseudo clusters with at least $S$ samples.

The level of label noise is a trade-off between the remaining data's size and the purifying process's intensity. With a more aggressive purifying process, the remaining training data's size becomes smaller, and the level of label noise is reduced. In the first few iterations, we apply an aggressive purifying process to the pseudo labels, which keeps 30\% to 50\% of training data. As the speaker model becomes more discriminative, we relax the purifying process and allow more data to train the network.

\subsection{Learning with Pseudo Labels}
After the pseudo labels purifying process, the remaining training dataset $\mathcal{D'}=\{\x_1, \cdots, \x_{N'}\}$ contains $N'$ utterances.
With the generated pseudo labels $\{y_1, \cdots, y_{N'}\}$, the speaker embedding network can be discriminatively trained with a parametrized classifier $g_W(\cdot)$ which predicts the labels for the speaker embedding $\z_i=f_\theta(\x_i)$. The parameters $\{\theta, W\}$ are jointly trained with the cross-entropy loss:
\begin{equation}
    \mathcal{L}_\mathrm{spk} = - \sum_{i=1}^{N'}\log \frac{\exp({g_{W}}_{y_i}(\z_i))}{\sum_{j=1}^{K}\exp({g_{W}}_{j}(\z_i))}
\end{equation}
where ${g_{W}}_{j}(\z_i)$ is the $j^\mathrm{th}$ element ($j\in[1, K]$) of the class score vector ${g_{W}}(\z_i)$, $K$ is the number of the pseudo class.

\subsection{Network Architecture}
We opt for a residual convolutional network (ResNet) to learn speaker representation from the spectral feature sequence of varying length \cite{cai_exploring_2018,He2016Deep}. The ResNet's output feature maps are aggregated with a global statistics pooling layer, which calculates means and standard deviations for each feature map. A fully connected layer is employed afterward to extract the 128-dimensional speaker embedding.

\section{Experiments}

\subsection{Dataset}
The experiments are conducted on the development set of Voxceleb 2, which contains 1,092,009 utterances from 5,994 speakers \cite{chung_voxceleb2:_2018}. Speaker labels are not used in the proposed method.

For evaluation, the development set and test set of Voxceleb 1 are used \cite{nagrani_voxceleb:_2017}. We report the speaker verification results on 3 trial sets as defined in \cite{chung_voxceleb2:_2018}: the \textit{original test set} of Voxceleb 1 containing 37,720 trials from 40 speakers, the \textit{Voxceleb 1-E} test set (using the entire dataset) containing 581,480 trials from 1251 speakers, the \textit{Voxceleb 1-H} test set (within the same nationality and gender) containing 552,536 trials from 1190 speakers.

\subsection{Data Augmentation}
Data augmentation is proven to be an effective strategy for both conventional learning with supervision \cite{cai_within-sample_2020} and contrastive self-supervision learning \cite{inoue_semi-supervised_2020,huh_augmentation_2020,chen_simple_2020} in the context of deep learning. We perform data augmentation with MUSAN dataset \cite{musan}. The noise type includes ambient noise, music, television, and babble noise for the background additive noise. The television noise is generated with one music file and one speech file. The babble noise is constructed by mixing three to eight speech files into one. For the reverberation, the convolution operation is performed with 40,000 simulated room impulse responses (RIR) in MUSAN. We only use RIRs from small and medium rooms.

With contrastive self-supervised learning, three augmentation types are randomly applied to each training utterance: applying only noise addition, applying only reverberation, and applying both noise and reverberation. The signal-to-noise ratios (SNR) are set between 5 to 20 dB.

When training with pseudo labels, either background noise or reverberation noise is added to the clean utterances 
with a probability of 0.6. The SNR is randomly set between 0 to 20 dB. 

\begin{table}[t]
    \caption{Clustering and label purifying results. $\{p, S\}$ are hyper-parameters defined in section 2.2.2. The number of remaining utterance after the label purifying process is presented. The normalized mutual information (NMI) values before and after the purifying process are also reported.}
    \label{tab: nmi}
    \centering
    \begin{tabular}{lcccc}
    \toprule 
   Model & $p$ & $S$ & \#utterances & NMI \cr
    \midrule
    CSL     & 0.6 & 8 & 347,625 & 0.8162 $\rightarrow$ 0.9381 \\
    Round 1 & 0.4 & 10 & 631,408 & 0.8669 $\rightarrow$ 0.9404 \\
    Round 2 & 0.4 & 10 & 644,692 & 0.8940 $\rightarrow$ 0.9603 \\
    Round 3 & 0.4 & 6 & 733,865 & 0.9114 $\rightarrow$ 0.9638 \\
    Round 4 & 0.3 & 6 & 843,770 & 0.9231 $\rightarrow$ 0.9618\\
    \bottomrule
    \end{tabular}
\end{table}

\begin{table*}[th]
  \caption{Speaker verification performance (minDCF and EER[\%]). The utterance and speaker number of the training data are presented. The NMIs of the pseudo labels for each iteration are also reported.}
  \label{tab:results}
  \centering
  \begin{tabular}[c]{lccccc|cc|cc}
    \toprule
    \textbf{Model} & \textbf{\#Utterances} & \textbf{\#Clusters} &\textbf{NMI} & \multicolumn{2}{c}{\textbf{Voxceleb 1 test}} & \multicolumn{2}{c}{\textbf{Voxceleb 1-E}} & \multicolumn{2}{c}{\textbf{Voxceleb 1-H}} \\
    \midrule
    Fully Supervised & 1,092,009 & 5,994 & 1 & 0.097 & 1.51 & 0.102 & 1.59 & 0.178 & 3.00 \\
    \midrule
    Nagrani \textit{et al.} \cite{nagrani_disentangled_2020} & 1,092,009 & - & - & - & 22.09 & - & - & - & - \\
    Chung \textit{et al.} \cite{chung_seeing_2020} & 1,092,009 & - & - & - & 17.52 & - & - & - & - \\
    Inoue \textit{et al.} \cite{inoue_semi-supervised_2020} & 1,092,009 & - & - & - & 15.26 & - & - & - & - \\
    Huh \textit{et al.} \cite{huh_augmentation_2020} & 1,092,009 & - & - & 0.454 & 8.65 & - & - & - & -\\
    \midrule
    Initial round (CSL) & 1,092,009 & - & - & 0.508 & 8.86 & 0.570 & 10.15 & 0.710 & 16.20 \\
    Round 1 & 347,625 & 2,839 & 0.9381 & 0.429 & 6.96 & 0.433 & 7.91 & 0.561 & 11.73 \\
    Round 2 & 631,408 & 4,776 & 0.9404 & 0.341 & 5.42 & 0.358 & 6.22 & 0.479 & 9.60 \\
    Round 3 & 644,692 & 4,708 & 0.9603 & 0.300 & 4.73 & 0.316 & 5.29 & 0.433 & 8.17 \\
    Round 4 & 733,865 & 5,018 & 0.9638 & 0.263 & 4.16 & 0.278 & 4.55 & 0.391 & 7.39 \\
    Round 5 & 843,770 & 5,407 & 0.9618 & 0.241 & 3.45 & 0.246 & 4.02 & 0.363 & 6.57 \\
    \bottomrule
  \end{tabular}
\end{table*}

\subsection{Implementation Details}

\subsubsection{Contrastive self-supervised learning setup}
For feature extraction, we choose a 40-dimensional log Mel-spectrogram with a 25ms Hamming window and 10ms shifts. The duration between 2 to 4 seconds is randomly generated for each data batch.

We use the same network architecture as in \cite{cai_within-sample_2020}. ReLU non-linear activation and batch normalization are applied to each convolutional layer in ResNet. Network parameters are updated using Adam optimizer \cite{kingma_adam_2017} with an initial learning rate of 0.001 and a batch size of 256. The temperature $\tau$ in equation (\ref{equation:csl}) is set as 0.1.

\subsubsection{Clustering setup}
The cluster number is set to 6,000 for \textit{k}-means. In Voxceleb 2, the audio segments are obtained with the self-supervised SyncNet \cite{chung_voxceleb2:_2018} from videos. We take advantage of this segment information and average the speaker embeddings from the same video. The \textit{k}-means clustering is performed on the averaged embeddings for the sake of clustering efficiency.

\subsubsection{Setup for models learned with pseudo labels}
For the network learned with pseudo labels, we use an 80-dimensional log Mel-spectrogram as input features. A duration between 3 to 4 seconds is randomly generated for each data batch. The network architecture is the same as the one used in CSL but with double feature map channels. Dropout is added before the speaker classification layer to prevent overfitting \cite{srivastava_dropout:_2014}. Network parameters are updated using stochastic gradient descent (SGD) algorithm. The learning rate is initially set to 0.1 and is divided by 10 whenever the training loss reaches a plateau.

\subsection{Experimental Results}
We use normalized mutual information (NMI) as a measurement of clustering quality. NMI measures the information shared between the true speaker labels $U$ and the predictive pseudo labels $V$. It is defined as:
\begin{equation}
	\mathrm{NMI}(U, V) = \frac{2 \times I(U; V)}{H(U) + H(V)}
\end{equation}
where $I(U;V)$ is the mutual information between $U$ and $V$, and $H(\cdot)$ denotes entropy. When two label assignments that are largely independent, NMI becomes 0. When they are in significant agreement, NMI equals to 1. Table \ref{tab: nmi} shows the NMI before and after the label purifying process for each round. We apply an aggressive purifying process in the first round and relax the process to allow more data to train the network in the following rounds. The original NMIs before the process increase as the iteration increase, which indicates the speaker embedding network becomes discriminative. Also, the increased NMI after the label purifying process demonstrates the effectiveness of the process.

For the speaker verification experiments, cosine similarity is used for scoring at the test stage. We use equal error rate (EER) and minimum detection cost function (minDCF) as the performance metric. The reported minDCF is evaluated with $P_{\mathrm{target}}=0.05, C_\mathrm{miss}=C_\mathrm{fa}=1$. Table \ref{tab:results} reports the experimental results. In round 1, with only 32\% of the training data and noisy pseudo labels, the speaker model outperforms the one trained with CSL by 21.4\% in terms of EER. Results in table \ref{tab:results} also show that the performance of the speaker verification system keeps improving with the increase of round number, which demonstrates the effectiveness of the proposed iterative self-supervised learning framework.  By comparing round 4 and round 5 to their previous rounds, we observed that the enlarged training data contributes to the performance improvements. We also see that although round 3 and round 2 has similar training data size (645k and 631k utterances), round 3 achieves a 12.8\% EER reduction comparing to round 2. This performance improvement comes from the improved clustering quality: the NMI is 0.94 for round 2's training data and increases to 0.96 for round 3.

\subsection{Discussion}
Since the pseudo labels generated in each round do not share cluster indexes, the speaker embedding network is trained from scratch in each round, and it takes longer time to train the model as training data increase. We argue that our proposed framework could benefit from the metric learning objectives, which assumes an open-set setting with unseen speakers \cite{chung2020defence}. With metric learning objectives, the speaker model at each round could continue learning from the last round with the newly generated pseudo labels.

To mitigate the adverse effects caused by the pseudo label noise, we use a simple label purifying process to exclude the data sample with little clustering confidence in this work. The more sophisticated solution may come from the deep learning solution overcoming label noise, such as curriculum loss \cite{lyu2019curriculum}, label smoothing \cite{pereyra2017regularizing}, and so on.

\section{Conclusion}
In this paper, we proposed an iterative framework for self-supervision speaker recognition. It consists of an initial round of contrastive self-supervised learning and several rounds of discriminative training with pseudo labels obtained by a clustering algorithm. The framework exploits DNN's ability to learn from data with noisy label and improves the model's discriminative power iteratively. Experimental results on Voxceleb show that our proposed framework improves speaker verification systems' performance compared to purely contrastive self-supervision learning.

\bibliographystyle{IEEEbib}
\bibliography{mybib}

\begin{thebibliography}{10}

\bibitem{snyder_x-vectors:_2018}
D.~Snyder, D.~Garcia-Romero, G.~Sell, D.~Povey, and S.~Khudanpur,
\newblock ``x-{vectors}: {Robust} {DNN} {Embeddings} for {Speaker}
  {Recognition},''
\newblock in {\em ICASSP}, 2018, pp. 5329--5333.

\bibitem{cai_exploring_2018}
W.~Cai, J.~Chen, and M.~Li,
\newblock ``Exploring the {Encoding} {Layer} and {Loss} {Function} in
  {End}-to-{End} {Speaker} and {Language} {Recognition} {System},''
\newblock in {\em Speaker Odyssey}, 2018, pp. 74--81.

\bibitem{chung2020defence}
J.~S. Chung, J.~Huh, S.~Mun, M.~Lee, H.~S. Heo, S.~Choe, C.~Ham, S.~Jung,
  B.~Lee, and I.~Han,
\newblock ``{In defence of metric learning for speaker recognition},''
\newblock in {\em {{Interspeech}}}, 2020.

\bibitem{colorization}
R.~Zhang, P.~Isola, and A.~Efros,
\newblock ``{Colorful Image Colorization},''
\newblock in {\em ECCV}, 2016, pp. 649--666.

\bibitem{latent2image}
A.~Radford, L.~Metz, and S.~Chintala,
\newblock ``{Unsupervised Representation Learning with Deep Convolutional
  Generative Adversarial Networks},''
\newblock {\em arXiv:1511.06434}, 2015.

\bibitem{cpc1}
A.~Oord, Y.~Li, and O.~Vinyals,
\newblock ``{Representation Learning with Contrastive Predictive Coding},''
\newblock {\em arXiv:1807.03748}, 2018.

\bibitem{moco}
K.~He, H.~Fan, Y.~Wu, S.~Xie, and R.~Girshick,
\newblock ``{Momentum Contrast for Unsupervised Visual Representation
  Learning},''
\newblock in {\em CVPR}, 2020, pp. 9729--9738.

\bibitem{chen_simple_2020}
T.~Chen, S.~Kornblith, M.~Norouzi, and G.~Hinton,
\newblock ``A {{Simple Framework}} for {{Contrastive Learning}} of {{Visual
  Representations}},''
\newblock {\em arXiv:2002.05709}, 2020.

\bibitem{stafylakis_self-supervised_2019}
T.~Stafylakis, J.~Rohdin, O.~Plchot, P.~Mizera, and L.~Burget,
\newblock ``{Self-Supervised Speaker Embeddings},''
\newblock in {\em Interspeech}, 2019, pp. 2863--2867.

\bibitem{ravanelli_learning_2019}
M.~Ravanelli and Y.~Bengio,
\newblock ``Learning {{Speaker Representations}} with {{Mutual Information}},''
\newblock in {\em Interspeech}, 2019, pp. 1153--1157.

\bibitem{inoue_semi-supervised_2020}
N.~Inoue and K.~Goto,
\newblock ``Semi-{{Supervised Contrastive Learning}} with {{Generalized
  Contrastive Loss}} and {{Its Application}} to {{Speaker Recognition}},''
\newblock {\em arXiv:2006.04326}, 2020.

\bibitem{huh_augmentation_2020}
J.~Huh, H.~S. Heo, J.~Kang, S.~Watanabe, and Joon~S. Chung,
\newblock ``{Augmentation Adversarial Training for Unsupervised Speaker
  Recognition},''
\newblock {\em arXiv:2007.12085}, 2020.

\bibitem{nagrani_disentangled_2020}
A.~Nagrani, J.~S. Chung, S.~Albanie, and A.~Zisserman,
\newblock ``Disentangled {{Speech Embeddings Using Cross}}-{{Modal
  Self}}-{{Supervision}},''
\newblock in {\em {{ICASSP}}}, 2020, pp. 6829--6833.

\bibitem{chung_seeing_2020}
S.~W. Chung, H.~G. Kang, and J.~S. Chung,
\newblock ``{Seeing Voices and Hearing Voices: Learning Discriminative
  Embeddings Using Cross-Modal Self-Supervision},''
\newblock {\em arXiv:2004.14326}, 2020.

\bibitem{song_learning_2020}
H.~Song, M.~Kim, D.~Park, and J.~Lee,
\newblock ``Learning from {{Noisy Labels}} with {{Deep Neural Networks}}: {{A
  Survey}},''
\newblock in {\em {{ECCV}}}, 2020.

\bibitem{liu_iterative_2014}
W.~Liu, Z.~Yu, and M.~Li,
\newblock ``An {{Iterative Framework}} for {{Unsupervised Learning}} in the
  {{PLDA}} based {{Speaker Verification}},''
\newblock in {\em ISCSLP}, 2014, pp. 78--82.

\bibitem{falcon_framework_2020}
W.~Falcon and K.~Cho,
\newblock ``A {{Framework For Contrastive Self}}-{{Supervised Learning And
  Designing A New Approach}},''
\newblock {\em arXiv:2009.00104}, 2020.

\bibitem{He2016Deep}
K.~He, X.~Zhang, S.~Ren, and J.~Sun,
\newblock ``{Deep Residual Learning for Image Recognition},''
\newblock in {\em CVPR}, 2016, pp. 770--778.

\bibitem{chung_voxceleb2:_2018}
J.~Son Chung, A.~Nagrani, and A.~Zisserman,
\newblock ``Voxceleb2: {{Deep Speaker Recognition}},''
\newblock in {\em Interspeech}, 2018, pp. 1086--1090.

\bibitem{nagrani_voxceleb:_2017}
A.~Nagrani, J.~S. Chung, and A.~Zisserman,
\newblock ``Voxceleb: {A} {Large}-{Scale} {Speaker} {Identification}
  {Dataset},''
\newblock in {\em Interspeech}, 2017, pp. 2616--2620.

\bibitem{cai_within-sample_2020}
D.~Cai, W.~Cai, and M.~Li,
\newblock ``Within-{Sample} {Variability}-{Invariant} {Loss} for {Robust}
  {Speaker} {Recognition} {Under} {Noisy} {Environments},''
\newblock in {\em {ICASSP}}, 2020, pp. 6469--6473.

\bibitem{musan}
D.~Snyder, G.~Chen, and D.~Povey,
\newblock ``{MUSAN}: {A} {Music}, {Speech}, and {Noise} {Corpus},''
\newblock {\em arXiv:1510.08484}, 2015.

\bibitem{kingma_adam_2017}
D.~P. Kingma and J.~Ba,
\newblock ``Adam: {{A Method}} for {{Stochastic Optimization}},''
\newblock in {\em {ICLR}}, 2015.

\bibitem{srivastava_dropout:_2014}
N.~Srivastava, G.~Hinton, A.~Krizhevsky, I.~Sutskever, and R.~Salakhutdinov,
\newblock ``Dropout: {{A Simple Way}} to {{Prevent Neural Networks}} from
  {{Overfitting}},''
\newblock {\em Journal of Machine Learning Research}, vol. 15, no. 1, pp.
  1929--1958, 2014.

\bibitem{lyu2019curriculum}
Y.~Lyu and I.~W. Tsang,
\newblock ``{Curriculum Loss: Robust Learning and Generalization Against Label
  Corruption},''
\newblock in {\em ICLR}, 2020.

\bibitem{pereyra2017regularizing}
G.~Pereyra, G.~Tucker, J.~Chorowski, {\L}.~Kaiser, and G.~Hinton,
\newblock ``Regularizing neural networks by penalizing confident output
  distributions,''
\newblock in {\em ICLR}, 2017.

\end{thebibliography}

\end{document}